\documentclass[conference]{IEEEtran}

\usepackage{amsfonts}
\usepackage{cite}
\usepackage{graphicx}
\usepackage{amssymb}
\usepackage{subfigure}
\usepackage{array}
\usepackage{color}
\usepackage{amsmath}
\usepackage{algorithmic}
\usepackage{algorithm}
\usepackage{cases}
\usepackage{bm}
\usepackage{mathrsfs}
\usepackage{epsfig}
\usepackage{psfrag}
\usepackage{url}
\usepackage{setspace}

\usepackage{etoolbox}
\makeatletter
\patchcmd{\@makecaption}
  {\scshape}
  {}
  {}
  {}
\makeatletter
\patchcmd{\@makecaption}
  {\\}
  {.\ }
  {}
  {}
\makeatother

\newtheorem{Thm}{Theorem}

\newtheorem{Prob}{Problem}

\newtheorem{Prop}{Proposition}

\IEEEoverridecommandlockouts

\begin{document}

\title{ML Estimation and MAP Estimation for Device Activities in Grant-Free Random Access with Interference}

\author{
\IEEEauthorblockN{Dongdong Jiang} \IEEEauthorblockA{Dept. of EE, Shanghai Jiao Tong University, China}
\and
\IEEEauthorblockN{Ying Cui}\IEEEauthorblockA{Dept. of EE, Shanghai Jiao Tong University, China}
\thanks{This work was supported in part by the National Key R\&D Program of China under Grant 2018YFB1801102.}
}

\maketitle

\begin{abstract}
Device activity detection is one main challenge in grant-free random access, which is recently proposed to support massive access for massive machine-type communications (mMTC). Existing solutions fail to consider interference generated by massive Internet of Things (IoT) devices, or important prior information on device activities and interference. In this paper, we consider device activity detection at an access point (AP) in the presence of interference generated by massive devices from other cells. 
We consider the joint maximum likelihood (ML) estimation and the joint maximum a posterior probability (MAP) estimation of both the device activities and interference powers, jointly utilizing tools from probability, stochastic geometry and optimization. Each
estimation problem is a difference of convex (DC) programming problem, and a coordinate descent algorithm is proposed to obtain a stationary point. The proposed ML estimation extends the existing ML estimation by considering the estimation of interference powers together with the estimation of device activities. The proposed MAP estimation further enhances the proposed ML estimation by exploiting prior distributions of device activities and interference powers. Numerical results show the substantial gains of the proposed joint estimation designs, and reveal the importance of explicit consideration of interference and the value of prior information in device activity detection.
\end{abstract}

\begin{IEEEkeywords}
Device activity detection, massive machine-type communications (mMTC), grant-free random access, maximum likelihood (ML) estimation, maximum a posterior probability (MAP) estimation. 
\end{IEEEkeywords}

\section{Introduction}

Driven by the proliferation of Internet of Things (IoT), 
massive machine-type communication (mMTC) has been identified 
as one of the three main use cases 
in the fifth generation (5G) cellular technologies \cite{Popovski17Mag}. 
Grant-free random access is an important technique for supporting mMTC. In grant-free random access, 
a main challenge is to 
identify the set of active devices in the presence of an excessive number of potential devices. 
Due to inherent sparse device activity in mMTC, device activity detection can be formulated as compressed sensing (CS) problems and solved by many CS-based algorithms.

In~\cite{Lau15ICC}, the authors consider device activity detection and channel estimation, 
and propose a modified Bayesian CS algorithm, which exploits the active device sparsity and chunk sparsity feature of the channel matrix. 
In~\cite{Bockelmann13}, the authors consider joint device activity and data detection, 
and apply the greedy group orthogonal matching pursuit (GOMP) algorithm, which exploits block-sparsity information of data. In \cite{Liu18TSP} and \cite{Chen18TSP}, the authors consider joint activity detection and channel estimation, 
and propose approximate message passing (AMP) algorithms.
Recently, maximum likelihood (ML) estimation-based device activity detection designs are proposed and analyzed in \cite{Caire18ISIT,Yu19ICC}. Specifically, in \cite{Caire18ISIT}, the authors formulate device activity detection as an ML estimation problem, in which the received signals at multiple antennas affect the detection results via their empirical covariance matrix.
In \cite{Yu19ICC}, the authors propose a covariance-based joint device activity and data detection scheme, and analyze the distribution of the estimation error in the massive MIMO regime. 

Note that \cite{Lau15ICC,Bockelmann13,Liu18TSP,Chen18TSP,Caire18ISIT,Yu19ICC} do not  consider interference generated by devices in other cells, and hence the resulting algorithms  may not provide desirable detection performance in practical mMTC with nonnegligible interference from massive IoT devices in other cells. In addition, notice that the ML-based algorithms in \cite{Caire18ISIT} and \cite{Yu19ICC} do not consider prior knowledge on sparsity patterns of device activities.
How to take into account the impact of interference and prior knowledge on device activities and interference powers to improve device activity detection remains an open problem. 

In this paper, we aim to tackle the above issues. We study activity detection for single-antenna devices at a multi-antenna access point (AP) in the presence of interference generated by massive devices from other cells. 
The main contributions of this paper are listed as follows.
\begin{itemize}
\item When prior distributions are not available, we consider the joint ML estimation of both device activities and interference powers. In particular, by carefully approximating interference powers, we first obtain a tractable expression for the likelihood of observations in the presence of interference. Then, we formulate the joint ML estimation problem which is a difference of convex (DC) programming problem. By making good use of the problem structure, we propose a coordinate descent algorithm which converges to a stationary point. The proposed ML estimation successfully extends the existing ML estimation~\cite{Caire18ISIT} to the practical scenario with interference. 


\item When prior distributions are known, we consider the joint MAP estimation of both device activities and interference powers. 
Using tools from stochastic geometry, we drive a tractable expression for the distributions of interference powers. Based on the prior distributions of device activities and interference powers together with the likelihood of observations, we formulate the joint MAP estimation problem which is also a DC programming problem. By exploiting the problem structure, we propose a coordinate descent algorithm which converges to a stationary point. The proposed MAP estimation further enhances the proposed ML estimation, by taking the prior information on device activities and interference powers into consideration. We also show that the influence of the prior information reduces as the number of antennas increases.


\item Finally, by numerical results, we show the substantial gains of the proposed designs over well-known existing designs. 
The numerical results also demonstrate the importance of explicit consideration of interference and the value of prior information in device activity detection.
\end{itemize}

\section{System Model}

\begin{figure}[t]
\begin{center}
 \includegraphics[width=5cm]{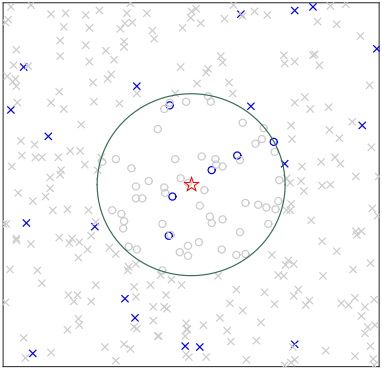}
  \end{center}
  \vspace{-2mm}
  \caption{\small{System model. The red star represents the AP, and the green circle represents its cell. The blue circles and crosses represent active devices inside and outside the cell, respectively. The gray circles and crosses represent inactive devices inside and outside the cell, respectively.}}
  \vspace{-4mm}
\label{fig:system_model}
\end{figure}

As shown in Fig.~\ref{fig:system_model}, we consider 
a large-scale wireless network which consists of $M$-antenna APs and single-antenna devices, and focus on device activity detection at a single AP in the presence of interference generated from massive active devices out of its cell. 
The devices are indexed by $i$ and the set of indices of devices is denoted as $\mathcal I\triangleq \{1,2,\cdots\}$. 
Let $a_i\in\{0,1\}$ denote the activity state of device $i$, where $a_i=1$ indicates that device $i$ is active and $a_i=0$ otherwise. 
Without loss of generality (w.l.o.g.), we assume that a typical AP is located at the origin. 
Let $\Phi_0$ denote the set of indices of the devices in the typical cell. Let $N\triangleq |\Phi_0|$ denote the number of devices in the typical cell. 
Denote $\mathbf a\triangleq (a_i)_{i\in\Phi_0}\in\{0,1\}^N$. 
Let $d_{i}$ denote the distance between device $i$ and the typical AP. 
For large-scale fading, consider power-law path loss model, i.e., transmitted signals with distance $d$ are attenuated with a factor $d^{-\alpha}$, where $\alpha\geq 2$ is the path loss exponent. Let $\gamma_{i}\triangleq d_{i}^{-\alpha}$ denote the path loss between device $i$ and the typical AP. Denote $\bm \gamma\triangleq (\gamma_{i})_{i\in\Phi_0}$. Assume that $\bm \gamma$ is perfectly known at the typical AP~\cite{Lau15ICC}. For small-scale fading, consider block fading channel model between devices and the typical AP, i.e., the state of a channel is static in each coherence block and changes across blocks. 
Let $\mathbf h_{i}\in\mathbb C^{M}$ denote the channel vector between device $i$ and the typical AP. We  assume Rayleigh fading and all $\mathbf h_{i}$, $i\in\mathcal I$ are independently and identically distributed (i.i.d.) according to $\mathcal{CN}(0,\mathbf I_M)$.

We consider a massive access scenario arising from mMTC, where each cell contains a large number of devices and very few of them are active in each coherence block.
We adopt a grant-free multiple-access scheme, 
where each device $i$ is assigned a specific pilot sequence $\mathbf p_i=(p_{i,\ell})_{\ell\in\mathcal L}\in \mathbb C^L$ of length $L$, which is much smaller than the number of devices in the typical cell.
Denote $\mathcal L\triangleq \{1,2,\cdots,L\}$. 
Let $\mathbf P\triangleq (\mathbf p_i)_{i\in\Phi_0}\in\mathbb C^{L\times N}$ denote the $L\times N$ matrix of the pilot sequences of the devices in the typical cell. 
As $L\ll N$, it is not possible to assign mutually orthogonal sequences to all $N$ devices. As in~\cite{Yu19ICC}, we assume that the pilot sequences for all devices (inside and outside the typical cell) are generated in an i.i.d. manner according to $\mathcal{CN}(0,\mathbf I_L)$.  Assume that $\mathbf P$ is known at the typical AP.
In each coherence block, all active devices synchronously send their pilot sequences to their associated APs, and
each AP detects the activities of its associated devices. 
Let $\mathbf Y\in\mathbb C^{L\times M}$ denote the received signal over $L$ signal dimensions and $M$ antennas at the typical AP. 
Then, we have 
\begin{align}
&\mathbf  Y = \mathbf P\mathbf A\bm \Gamma^{\frac{1}{2}}\mathbf H^T+\sum_{i\in\Phi}a_id_{i}^{-\frac{\alpha}{2}}\mathbf p_i \mathbf h_{i}^T+\mathbf Z, \label{eqn:receive_signal}
\end{align}
where $\Phi\triangleq \mathcal I\setminus\Phi_0$, 
$\mathbf H\triangleq (\mathbf h_i)_{i\in\Phi_0}\in\mathbb C^{M\times N}$, 
$\mathbf \Gamma\triangleq{\rm diag}(\bm\gamma)$, $\mathbf A\triangleq{\rm diag}(\mathbf a)$, and $\mathbf Z$ is the additive white Gaussian noise (AWGN) with each element following $\mathcal{CN}(0,\mathbf \delta^2)$. Note that the first term in \eqref{eqn:receive_signal} is the received signal from the devices in the typical cell and the second term is the received interference from the devices out of the typical cell.




\section{Joint ML Estimation of Device Activities and Interference Powers}

\subsection{Likelihood Function}
Let $\mathbf y_{m}$ denote the $m$-th column of $\mathbf Y$.  Under Rayleigh fading and AWGN, $\mathbf y_m$, $m\in\{1,2,\cdots,M\}$ 
are i.i.d. according to $\mathcal{CN}(0,\mathbf P\mathbf A\mathbf \Gamma\mathbf P^H+\widetilde{\mathbf X}+\delta^2\mathbf I_L)$, where $\widetilde{\mathbf X}\triangleq \sum_{i\in \Phi}a_i\gamma_{i}\mathbf p_i\mathbf p_i^H\in\mathbb C^{L\times L}$.
Here, $\mathbf P\mathbf A\mathbf \Gamma\mathbf P^H$, $\widetilde{\mathbf X}$ and $\delta^2\mathbf I_L$ are the covariance matrices of the received signal, interference and noise at the typical AP, respectively. Besides $\mathbf a$, $\widetilde{\mathbf X}$ is also unknown and has to be estimated. Notice that the estimation of the $L\times L$ matrix $\widetilde{\mathbf X}$ involves high complexity, especially when $L$ is moderate or large. In addition, note that 
$\widetilde{\mathbf X}$ is diagonally dominant, as pilot sequences are generated from i.i.d. complexed Gaussian distribution~\cite{Yu19ICC}. Therefore, we approximate
$\widetilde{\mathbf X}$ with $\mathbf X\triangleq  {\rm diag}(\mathbf x)$, where $\mathbf x=(x_\ell)_{\ell\in\mathcal L}\in[0,\infty)^L$. Later, we shall see that allowing entries of $\mathbf x$ to be different facilitates the coordinate descent optimization in device activity detection and also provides reasonable detection accuracy. Note that $\mathbf x$ can be interpreted as the interference powers at the $L$ signal dimensions. In addition, rewriting $\mathbf X$ as $\sum_{\ell\in\mathcal L}x_{\ell}\mathbf e_\ell\mathbf e_\ell^H$, where $\mathbf e_\ell$ is the $\ell$-th standard basis which has a $1$ as its $\ell$-th entry and $0$s elsewhere, the interference can be viewed as from $L$ active devices with pilots $\mathbf e_\ell$, $\ell\in\mathcal L$ and pass losses $x_\ell$, $\ell\in\mathcal L$.
Under the approximation of $\widetilde{\mathbf X}$, the distribution of $\mathbf y_{m}$ is given by
\begin{align}
\mathbf y_{m}\sim \mathcal{CN}\left(0, \mathbf P\mathbf A\bm \Gamma\mathbf P^H+\mathbf X+\delta^2\mathbf I_L\right).\label{eqn:distribution_Y}
\end{align}
Based on \eqref{eqn:distribution_Y} and the fact that $\mathbf y_{m}$, $m\in\{1,2,\cdots, M\}$ are i.i.d. vectors, the likelihood function of $\mathbf  Y$ is given by
\begin{align}
&f(\mathbf Y|\mathbf a,\mathbf  x)\notag\\
&\propto \frac{\exp\left(-{\rm tr}\left(\left(\mathbf P\mathbf A\mathbf \Gamma\mathbf P^H+\mathbf X+\delta^2\mathbf I_L\right)^{-1}\mathbf Y\mathbf Y^H\right)\right)}{\vert(\mathbf P\mathbf A\mathbf \Gamma\mathbf P^H+\mathbf X+\delta^2\mathbf I_L)\vert^M},\notag
\end{align}
where $\propto$ means ``proportional to'', $\vert\cdot\vert$ is the determinant of a matrix, and ${\rm tr}(\cdot)$ is the trace of a matrix. Note that the constant coefficient is omitted for notation simplicity.

\subsection{ML Estimation}

In this part, we perform 
the joint ML estimation of $N$ device activities $\mathbf a$ and $L$ interference powers $\mathbf x$.
The maximization of the likelihood function $f(\mathbf Y|\mathbf a,\mathbf  x)$ is equivalent to the minimization of the negative log-likelihood function $-\log f(\mathbf Y|\mathbf a,\mathbf  x)\propto f_{\rm ML}(\mathbf a, \mathbf x)$, where
\begin{align}
f_{\rm ML}(\mathbf a, \mathbf x)
\triangleq & \log|\mathbf P\mathbf A\mathbf \Gamma\mathbf P^H+\mathbf X+\delta^2\mathbf I_L|\notag\\
&+{\rm tr}((\mathbf P\mathbf A\mathbf \Gamma\mathbf P^H+\mathbf X +\delta^2\mathbf I_L)^{-1}\widehat{\mathbf \Sigma}_{\mathbf Y}), \notag
\end{align}
with 
$\widehat{\mathbf \Sigma}_{\mathbf Y} \triangleq \frac{1}{M}\mathbf Y\mathbf Y^H$. 
By omitting the constant term in the negative log-likelihood function, the joint ML estimation problem can be simplified to:\footnote{In this paper, as in~\cite{Caire18ISIT} and \cite{Yu19ICC}, binary condition $a_i\in\{0,1\}$ is relaxed to $a_i\in [0,1]$ in estimation problems, and activity detection is conducted by performing thresholding on the solutions of the relaxed problems.}
\begin{Prob}[Joint ML Estimation]\label{Prob:ML}
\begin{align}
\min_{\mathbf a,\mathbf x} &\quad f_{\rm ML}(\mathbf a, \mathbf x)\notag\\
s.t. &\quad  1\geq a_i\geq 0,\quad i\in \Phi_0,\notag\\
&\quad x_\ell\geq 0,\quad \ell\in\mathcal L.\notag
\end{align}
Let $(\mathbf a^*,\mathbf x^*)$ denote an optimal solution of Problem~\ref{Prob:ML}.
\end{Prob}

Different from the ML estimation problems in \cite{Caire18ISIT} and \cite{Yu19ICC}, which focus only on estimating $\mathbf a$ in a single-cell network without interference, Problem~\ref{Prob:ML} considers the joint estimation of $\mathbf a$ and $\mathbf x$ in the presence of interference.
Due to the existence of the extra nonnegative diagonal matrix $\mathbf X$ in $f_{\rm ML}(\mathbf a, \mathbf x)$, $\mathbf a^*$ obtained from the proposed joint ML estimation is likely to have fewer zero elements than the optimal solution of the ML estimation in \cite{Caire18ISIT}, when $\mathbf Y$ contains interference.

As $\log|\mathbf P\mathbf A\mathbf \Gamma\mathbf P^H+\mathbf X+\delta^2\mathbf I_L|$ is a concave function of $\mathbf a$ and $\mathbf x$, and ${\rm tr}((\mathbf P\mathbf A\mathbf \Gamma\mathbf P^H+\mathbf X+\delta^2\mathbf I_L)^{-1}\widehat{\mathbf \Sigma}_{\mathbf Y})$ is a convex function of $\mathbf a$ and $\mathbf x$,  
$f_{\rm ML}(\mathbf a, \mathbf x)$ is a difference of convex (DC) function. In addition, by noting that the inequality constraints are linear, Problem~\ref{Prob:ML} is a DC programming problem, which is a subcategory of non-convex problems. Note that obtaining a stationary point is the classic goal for solving a non-convex problem.
In the following, we extend the coordinate descend method in~\cite{Caire18ISIT} for the case without interference to obtain a stationary point of Problem~\ref{Prob:ML} for the case with interference.  As a closed-form optimal solution can be obtained
for the optimization of each coordinate, the coordinate descent method is more computationally efficient than standard methods for DC programming, such as convex-concave procedure.

At each step of the proposed coordinate descent algorithm, we optimize $f_{\rm ML}(\mathbf a,\mathbf x)$ with respect to one of the coordinates in $\{a_i$: $i\in\Phi_0\}\cup\{x_\ell:\ell\in\mathcal L\}$. 
Specifically, given $\mathbf a$ and $\mathbf x$ obtained in the previous step, the coordinate descend optimization with respect to $a_i$, $i\in\Phi_0$ is equivalent to the optimization of the increment $d$ in $a_i$:
\begin{align}
\min_{1-a_i\geq d\geq -a_i} f_{\rm ML}(\mathbf a + d\mathbf e_i, \mathbf x),\label{eqn:ML_a}
\end{align}
and the coordinate descend optimization with respect to $x_\ell$, $\ell\in\mathcal L$ is equivalent to the optimization of the increment $d$ in $x_\ell$:
\begin{align}
\min_{d\geq -x_\ell} f_{\rm ML}(\mathbf a ,\mathbf x+ d\mathbf e_\ell).\label{eqn:ML_x}
\end{align}

Based on structural properties of the coordinate descent optimization problems in~\eqref{eqn:ML_a} and \eqref{eqn:ML_x}, we can derive their closed-form optimal solutions.
\begin{Thm}[Optimal Solutions of Coordinate Descent Optimizations in \eqref{eqn:ML_a} and \eqref{eqn:ML_x}]\label{Thm:Step_one_AP}
Given $\mathbf a$ and $\mathbf x$ obtained in the previous step,
the optimal solution of the coordinate optimization with respect to $a_i$ in \eqref{eqn:ML_a} is given by
\begin{align}
\min\left\{\max\left\{\frac{\mathbf p_{i}^H\mathbf \Sigma^{-1}\widehat{\mathbf \Sigma}_{\mathbf Y}\mathbf \Sigma^{-1}\mathbf p_{i}-\mathbf p_{i}^H\mathbf \Sigma^{-1}\mathbf p_{i}}{\gamma_{i}(\mathbf p_{i}^H\mathbf \Sigma^{-1}\mathbf p_{i})^2},-a_{i}\right\},1-a_i\right\},\label{eqn:d_ML_a}
\end{align}
and the optimal solution of the coordinate optimization with respect to $x_\ell$ in \eqref{eqn:ML_x} is given by
\begin{align}
\max\left\{\frac{\mathbf e_{\ell}^H\mathbf \Sigma^{-1}\widehat{\mathbf \Sigma}_{\mathbf Y}\mathbf \Sigma^{-1}\mathbf e_{\ell}-\mathbf e_{\ell}^H\mathbf \Sigma^{-1}\mathbf e_{\ell}}{(\mathbf e_{\ell}^H\mathbf \Sigma^{-1}\mathbf e_{\ell})^2},-x_{\ell}\right\}.\label{eqn:d_ML_x}
\end{align}
Here, $\mathbf \Sigma \triangleq \mathbf P\mathbf A\bm \Gamma\mathbf P^H+\mathbf X+\delta^2\mathbf I_L$ is determined by $\mathbf a$ and $\mathbf x$.
\end{Thm}

The details of the coordinate descent algorithm for solving Problem~\ref{Prob:ML} are summarized in Algorithm~1. Specifically, in Steps $4-5$,  each coordinate of $\mathbf a$ is updated. In Steps $8-9$, each coordinate of $\mathbf x$ is updated. Different from the coordinate optimization in \cite{Caire18ISIT} which only updates the coordinates of $\mathbf a$, the coordinate updates in Algorithm~\ref{alg:ML_descend} are with respect to both $\mathbf a$ and $\mathbf x$. In addition, as in~\cite{Yu19ICC}, we update $\mathbf \Sigma^{-1}$ rather than $\mathbf \Sigma$ in each coordinate descend optimization (i.e., Steps $5$ and $9$), which avoids the calculation of matrix inversion and improves the computation efficiency of the algorithm. As the coordinate descent optimizations are solved optimally, we can obtain a
stationary point of Problem~\ref{Prob:ML} using Algorithm~\ref{alg:ML_descend}.

\begin{algorithm} \caption{Coordinate Descend Algorithm for Joint ML Estimation}
\small{\begin{algorithmic}[1]
\STATE Initialize $\bm \Sigma^{-1}=\frac{1}{\delta^2} \mathbf I_L$, $\mathbf a=\mathbf 0$, $\mathbf x=\mathbf 0$.
\STATE \textbf{repeat}
\FOR {$i\in\Phi_0$}
\STATE Calculate $d$ according to \eqref{eqn:d_ML_a}. 
\STATE Update $a_{i}=a_{i}+d$ and $\mathbf \Sigma^{-1} = \mathbf \Sigma^{-1}-\frac{d\gamma_{i}\mathbf \Sigma^{-1}\mathbf p_{i}\mathbf p_{i}^H\mathbf \Sigma^{-1}}{1+d\gamma_{i}\mathbf p_{i}^H\mathbf \Sigma^{-1}\mathbf p_{i}}$.
\ENDFOR
\FOR {$\ell\in\mathcal L$}
\STATE Calculate $d$ according to \eqref{eqn:d_ML_x}. 
\STATE Update $x_{\ell}=x_{\ell}+d$ and $\mathbf \Sigma^{-1} = \mathbf \Sigma^{-1}-\frac{d\mathbf \Sigma^{-1}\mathbf e_{\ell}\mathbf e_{\ell}^H\mathbf \Sigma^{-1}}{1+d\mathbf e_{\ell}^H\mathbf \Sigma^{-1}\mathbf e_{\ell}}$.
\ENDFOR
\STATE \textbf{until} $(\mathbf a,\mathbf x)$ satisfies some stopping criterion.
\end{algorithmic}}\normalsize\label{alg:ML_descend}
\end{algorithm}

\section{Joint MAP Estimation of Device Activities and Interference Powers}

In this section, we assume that $\mathbf a$ and $\mathbf x$ are random, and we perform the joint MAP estimation of $\mathbf a$ and $\mathbf x$. 

\subsection{Prior Distributions}

We assume that $\mathbf a$ and $\mathbf x$ are independently distributed. Note that this is a weak assumption, as it only requires that the device activities in the typical cell are independent of those in the other cells. 
First, we introduce a prior distribution of the Bernoulli random vector $\mathbf a$. For tractability, we assume that devices access the channel with probability $p_a\ll 1$ in an i.i.d. manner. The probability mass function (p.m.f.) of $\mathbf a$ is given by
\begin{align}
p(\mathbf a)=\exp\left(\log \frac{p_a}{1-p_a}\sum_{i\in\Phi_0}a_i+N\log(1-p_a)\right).\notag
\end{align}

Next, we derive a prior distribution of $\mathbf x$. For ease of analysis, as in~\cite{Chen18TSP}, we assume that the coverage area of the typical AP is a disk with radius $R$, and the devices associated with the other APs out of the typical cell are distributed according to a homogeneous PPP with density $\lambda$, which is a widely adopted model for large-scale wireless
networks. As pilot sequences are generated from i.i.d. complexed Gaussian distribution, the diagonal entries of $\widetilde{\mathbf X}$ are i.i.d.. We assume that $x_{\ell}$, $\ell\in\mathcal L$ are i.i.d. with the same distribution as $\sum_{i\in\Phi} a_i\gamma_{i}$. 
Therefore, $x_\ell$ is a power-law shot noise, whose exact distribution is still not known. As in~\cite{Aljuaid10TVT}, we use a Gaussian distribution to approximate the probability density function (p.d.f.) of $x_\ell$. Note that the Gaussian approximation is accurate when the cell size, i.e., $R$  is large
\cite{Aljuaid10TVT,Hasan07TWC}. 
Based on the above assumptions and techniques from stochastic geometry, we have the following results.
\begin{Prop}[Approximate Distribution of $\mathbf x$]\label{Lem:distribution_X}
The p.d.f. of $\mathbf x$ is given by 
\begin{align}
g(\mathbf x) \approx \frac{1}{(\sqrt{2\pi}\sigma)^L}\exp\left(-\frac{\sum_{\ell\in\mathcal L}(x_\ell-\mu)^2}{2\sigma^2}\right),\notag
\end{align}
where 
$\mu \triangleq\mathbb E(x_\ell)= \frac{2\pi\lambda p_a R^{2-\alpha}}{\alpha-2}$, 
$\sigma^2 \triangleq {\rm Var}(x_\ell)=\frac{\pi\lambda p_a R^{2-2\alpha}}{\alpha-1}$.
\end{Prop}

\begin{figure}[t]
\begin{center}
 \includegraphics[width=8cm]{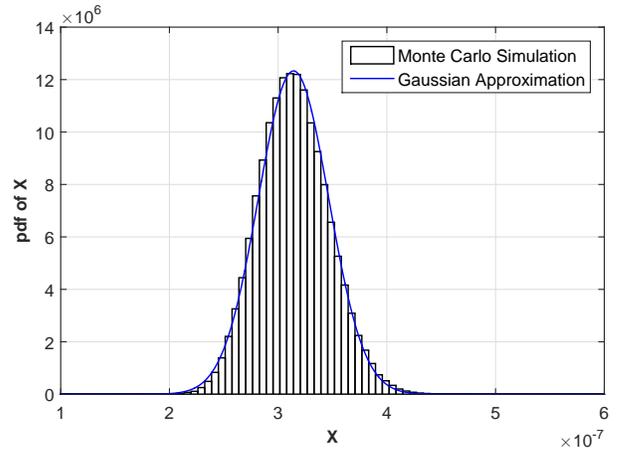}
  \end{center}
  \caption{\small{Comparison between the p.d.f. of $x_\ell$ and its corresponding Gaussian approximation. $R=100$, $\lambda=0.01$, $p_a=0.1$ and $\alpha=4$.}}
\label{fig:gamma_approximation}
\end{figure}

Fig.~\ref{fig:gamma_approximation} plots the histogram of $x_\ell$, which reflects the shape of the p.d.f. of $x_\ell$, as well as the Gaussian distribution with the same mean and variance. From Fig.~\ref{fig:gamma_approximation}, we can see that the Gaussian distribution is a good approximation of the exact p.d.f. of $x_\ell$ under the considered simulation setup, which verifies Proposition~\ref{Lem:distribution_X}.

\subsection{MAP Estimation}

Based on the likelihood function of $\mathbf Y$ and the prior distributions of $\mathbf a$ and $\mathbf x$, the joint posterior distribution of $\mathbf a$ and $\mathbf x$, given the observation $\mathbf Y$, is given by
\begin{align}
&f(\mathbf a,\mathbf  x|\mathbf Y) \propto f(\mathbf Y|\mathbf a,\mathbf  x)p(\mathbf a)g(\mathbf x)\notag\\
&\propto \frac{\exp\left(-{\rm tr}\left(\left(\mathbf P\mathbf A\mathbf \Gamma\mathbf P^H+\mathbf X+\delta^2\mathbf I_L\right)^{-1}\mathbf Y\mathbf Y^H\right)\right)}{\vert(\mathbf P\mathbf A\bm \Gamma\mathbf P^H+\mathbf X+\delta^2\mathbf I_L)\vert^M}\notag\\
&\quad\times\exp\left(-\sum_{\ell\in\mathcal L}\frac{(x_\ell-\mu)^2}{2\sigma^2}+\ln \frac{p_a}{1-p_a}\sum_{i\in\Phi_0}a_i\right).\notag
\end{align}
The maximization of the posterior probability $f(\mathbf a,\mathbf  x|\mathbf Y)$ is equivalent to the minimization of the negative logarithm of the posterior probability $-\log f(\mathbf a,\mathbf  x|\mathbf Y)\propto f_{\rm MAP}(\mathbf a,\mathbf x)$, where
\begin{align}
f_{\rm MAP}(\mathbf a,\mathbf x)\triangleq & \frac{1}{2M\sigma^2}\sum_{\ell\in\mathcal L}(x_\ell-\mu)^2-\frac{1}{M}\log \frac{p_a}{1-p_a}\sum_{i\in\Phi_0}a_i\notag\\
&+f_{\rm ML}(\mathbf a, \mathbf x).\notag
\end{align}
Note that $\frac{1}{2M\sigma^2}\sum_{\ell\in\mathcal L}(x_\ell-\mu)^2$ is from the p.d.f. of $\mathbf x$ and $-\frac{1}{M}\log \frac{p_a}{1-p_a}\sum_{i\in\Phi_0}a_i$ is from the p.m.f. of $\mathbf a$.  
The joint MAP estimation of $\mathbf a$ and $\mathbf x$ can be formulated as follows.
\begin{Prob}[Joint MAP Estimation]\label{Prob:MAP}
\begin{align}
\min_{\mathbf a,\mathbf x} &\quad f_{\rm MAP}(\mathbf a,\mathbf x)\notag\\
s.t. &\quad 1\geq a_i\geq 0,\quad i\in\Phi_0,\notag\\
&\quad x_\ell \geq 0,\quad \ell\in\mathcal L.\notag
\end{align}
Let $(\mathbf a^{\dagger},\mathbf x^\dagger)$ denote an optimal solution of Problem~\ref{Prob:MAP}.
\end{Prob}

By comparing $f_{\rm MAP}(\mathbf a,\mathbf x)$ with $f_{\rm ML}(\mathbf a,\mathbf x)$, we can draw the following conclusions. The incorporation of prior distribution $g(\mathbf x)$ pushes the estimate of $x_{\ell}$ towards its mean $\mu$ for all $\ell\in\mathcal L$. 
Since $p_a\ll 1$, the incorporation of the prior distribution $p(\mathbf a)$ pushes the estimate of $a_i$ towards $0$. As $f_{\rm MAP}(\mathbf a,\mathbf x)-f_{\rm ML}(\mathbf a,\mathbf x)$ decreases with $M$, the impacts of the prior distributions of $\mathbf a$ and $\mathbf x$ reduce as $M$ increases. As $M\to \infty$, $f_{\rm MAP}(\mathbf a,\mathbf x)\to f_{\rm ML}(\mathbf a,\mathbf x)$, Problem~\ref{Prob:MAP} reduces to Problem~\ref{Prob:ML}, and $(\mathbf a^\dagger,\mathbf x^\dagger)$  becomes $(\mathbf a^*,\mathbf x^*)$.

Similarly, we can see that $f_{\rm MAP}(\mathbf a,\mathbf x)$ is a DC function and Problem~\ref{Prob:MAP} is a DC programming problem.
We adopt the coordinate descend method to obtain a stationary point of Problem~\ref{Prob:MAP}. 
Specifically, given $\mathbf a$ and $\mathbf x$ obtained in the previous step, the coordinate descend optimization with respect to $a_i$, $i\in\Phi_0$ is given by 
\begin{align}
\min_{1-a_i\geq d\geq -a_i} \ f_{\rm MAP}(\mathbf a + d\mathbf e_i, \mathbf x),\label{eqn:MAP_a}
\end{align}
and the coordinate descend optimization with respect to $x_\ell$, $\ell\in\mathcal L$ is given by 
\begin{align}
\min_{d\geq -x_\ell} \ f_{\rm MAP}(\mathbf a ,\mathbf x+ d\mathbf e_\ell).\label{eqn:MAP_x}
\end{align}
Define 
\begin{align}
f_{x,\ell}(d,\mathbf a,\mathbf x)\triangleq &
\frac{(x_\ell-\mu+d)^2}{2M\sigma^2}
-\frac{d\mathbf e_\ell^H\mathbf \Sigma^{-1}\widehat{\mathbf \Sigma}_{\mathbf Y}\mathbf \Sigma^{-1}\mathbf e_\ell}{1+d\mathbf e_\ell^H\mathbf \Sigma^{-1}\mathbf e_\ell}\notag\\
&+\log(1+d\mathbf e_\ell^H\mathbf \Sigma^{-1}\mathbf e_\ell),\notag\\
h_{x,\ell}(d,\mathbf a,\mathbf x)\triangleq &
\frac{d+x_\ell-\mu}{M\sigma^2}
-\frac{\mathbf e_\ell^H\mathbf \Sigma^{-1}\widehat{\mathbf \Sigma}_{\mathbf Y}\mathbf \Sigma^{-1}\mathbf e_\ell}{(1+d\mathbf e_\ell^H\mathbf \Sigma^{-1}\mathbf e_\ell)^2}\notag\\
&+\frac{\mathbf e_\ell^H\mathbf \Sigma^{-1}\mathbf e_\ell}{1+d\mathbf e_\ell^H\mathbf \Sigma^{-1}\mathbf e_\ell}.\notag
\end{align}
We write $f_{x,\ell}(d,\mathbf a,\mathbf x)$ and $h_{x,\ell}(d,\mathbf a,\mathbf x)$ as functions of $\mathbf a$ and $\mathbf x$, as $\mathbf \Sigma$ is a function of $\mathbf a$ and $\mathbf x$. 
Note that $h_{x,\ell}(d,\mathbf a,\mathbf x)$ is the derivative function of $f_{x,\ell}(d,\mathbf a,\mathbf x)$ with respect to $d$. 
Denote $\mathcal X_{\ell}(\mathbf a,\mathbf x)\triangleq \{d\geq-x_\ell:h_{x,\ell}(d,\mathbf a,\mathbf x)=0\}$ as the set of roots of equation $h_{x,\ell}(d,\mathbf a,\mathbf x)=0$ that are no smaller than $-x_\ell$. 
Based on structural properties of the coordinate optimization problems, we have the following results.
\begin{Thm}[Optimal Solutions of Coordinate Descent Optimizations in~\eqref{eqn:MAP_a} and \eqref{eqn:MAP_x}]\label{Thm:Step_APs}
Given $\mathbf a$ and $\mathbf x$ obtained in the previous step, 
the optimal solution of the coordinate optimization with respect to $a_i$ in~\eqref{eqn:MAP_a} 
is given by \eqref{eqn:d_MAP_a}, as shown at the top of the next page,
\begin{figure*}
\begin{align}
\min\Bigg\{&\max\Bigg\{\frac{M}{2\log(\frac{p_a}{1-p_a})}\Bigg(1-\sqrt{1-\frac{\frac{4\gamma_{i}}{M}\log(\frac{p_a}{1-p_a})\mathbf p_i^H\mathbf \Sigma^{-1}\widehat{\mathbf \Sigma}_{\mathbf Y}\mathbf \Sigma^{-1}\mathbf p_i}{(\gamma_{i}\mathbf p_i^H\mathbf \Sigma^{-1}\mathbf p_i)^2}}\Bigg)-\frac{1}{\gamma_{i}\mathbf p_i^H\mathbf \Sigma^{-1}\mathbf p_i},-a_i\Bigg\}, 1-a_i\Bigg\}\label{eqn:d_MAP_a}
\end{align}
\end{figure*}
 and the optimal solution of the coordinate optimization with respect to $x_\ell$ in \eqref{eqn:MAP_x} 
is given by 
\begin{align}
\mathop{\arg\min}\limits_{d\in\mathcal X_\ell(\mathbf a,\mathbf x)\cup\{-x_\ell\}}f_{x,\ell}(d,\mathbf a,\mathbf x).\label{eqn:d_MAP_x}
\end{align}
\end{Thm}

From Theorem \ref{Thm:Step_APs}, we can see that in the coordinate descend optimizations, prior information on $\mathbf a$ and $\mathbf x$ affects the updates of $a_i$, $i\in\Phi_0$ and $x_\ell$, $\ell\in\mathcal L$, respectively. The roots of equation $h_{x,\ell}(d,\mathbf a,\mathbf x)=0$ can be easily obtained by solving a cubic equation with one variable, which has closed-form solutions. Thus, the coordinate descent optimizations can be efficiently solved.
The details of the coordinate descent algorithm for solving Problem~\ref{Prob:MAP} are summarized in Algorithm~\ref{alg:MAP_descend}.
Similarly, it is obvious that we can obtain a stationary point of Problem~\ref{Prob:MAP} by Algorithm~\ref{alg:MAP_descend}.

\begin{algorithm} \caption{Coordinate Descent Algorithm for Joint MAP Estimation}
\small{\begin{algorithmic}[1]
\STATE Initialize $\bm \Sigma^{-1}=\frac{1}{\delta^2} \mathbf I_L$, $\mathbf a=\mathbf 0$, $\mathbf x=\mathbf 0$.
\STATE \textbf{repeat}
\FOR {$i\in\Phi_0$}
\STATE Calculate $d$ according to \eqref{eqn:d_MAP_a}.
\STATE Update $a_{i}=a_{i}+d$ and $\mathbf \Sigma^{-1} = \mathbf \Sigma^{-1}-\frac{d\gamma_{i}\mathbf \Sigma^{-1}\mathbf p_{i}\mathbf p_{i}^H\mathbf \Sigma^{-1}}{1+d\gamma_{i}\mathbf p_{i}^H\mathbf \Sigma^{-1}\mathbf p_{i}}$.
\ENDFOR
\FOR {$\ell\in\mathcal L$}
\STATE Calculate $d$ according to \eqref{eqn:d_MAP_x}.
\STATE Update $x_{\ell}=x_{\ell}+d$ and $\mathbf \Sigma^{-1} = \mathbf \Sigma^{-1}-\frac{d\mathbf \Sigma^{-1}\mathbf e_{\ell}\mathbf e_{\ell}^H\mathbf \Sigma^{-1}}{1+d\mathbf e_{\ell}^H\mathbf \Sigma^{-1}\mathbf e_{\ell}}$.
\ENDFOR
\STATE \textbf{until} $(\mathbf a,\mathbf x)$ satisfies some stopping criterion.
\end{algorithmic}}\normalsize\label{alg:MAP_descend}
\end{algorithm}

\section{Numerical Results}

\begin{figure}
\begin{center}
\subfigure[\small{Density of interfering devices $\lambda$.}]
{\resizebox{7.5cm}{!}{\includegraphics{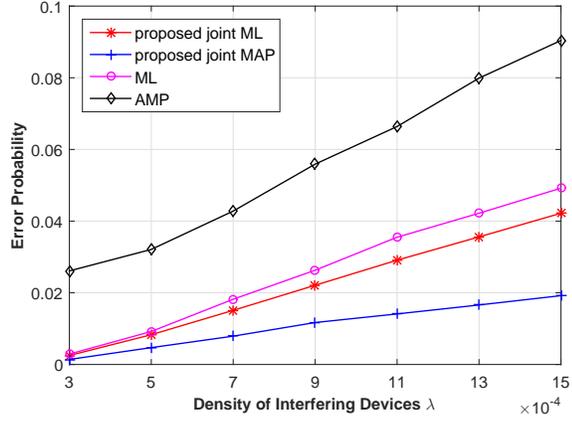}}}\quad
\subfigure[\small{Length of pilot sequences $L$.}]
{\resizebox{7.6cm}{!}{\includegraphics{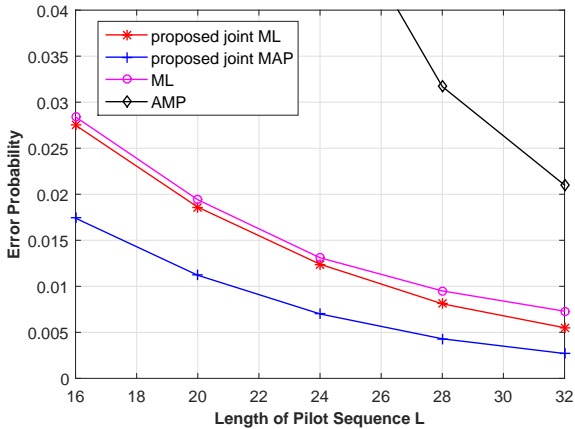}}}\quad
\subfigure[\small{Number of antennas $M$.}]
{\resizebox{7.6cm}{!}{\includegraphics{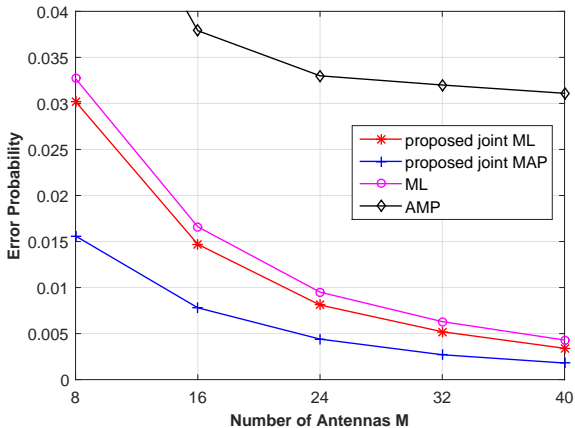}}}
\end{center}
\caption{\small{Error probability versus density of interfering devices $\lambda$, number of antennas $M$ and length of pilot sequences $L$.}}
\label{fig:performance_comparison}
\end{figure}

In this section, we evaluate the performance of the proposed activity detection designs via numerical results.
We compare the proposed designs with two existing designs, i.e., AMP in~\cite{Liu18TSP} and ML in~\cite{Caire18ISIT}, which do not consider interference. 
In the simulation, we consider that the typical cell is a disk with radius $R$, the $N$ devices in $\Phi_0$ are uniformly and randomly distributed in the typical cell, the devices associated with the other APs out of the typical cell are distributed according to a homogeneous PPP with density $\lambda$, and all devices access the channel with probability $p_a$ in an i.i.d. manner. 
We independently generate $2000$ realizations for the locations of devices, $a_i$ and $\mathbf h_i$, $i\in\mathcal I$, perform device activity detection in each realization, and evaluate the average error probability over all $2000$ realizations.
For the proposed designs and ML,
let $\hat{a}_i\triangleq \mathbf 1[\gamma_i\geq \theta]$ denote the estimate of the activity state of device $i\in \Phi_0$, where $\mathbf 1[\cdot]$ is the indicator function and $\theta>0$ is a threshold.
For AMP, let $\hat{a}_i\triangleq \mathbf 1[{\rm LLR}_i\geq 0]$ denote the estimate of the activity state of device $i$ \cite{Liu18TSP}, where ${\rm LLR}_i$ is the log-likelihood ratio for device $i$ and is given by (30) in~\cite{Liu18TSP}.
A detection error happens when $\hat{a}_i\neq a_i$.
For each of the proposed designs and ML, we evaluate the average error probability for $\theta\in\{0.01,0.02,\cdots,3\}$ and choose the minimum one as its average error probability.
In the simulation, unless otherwise stated, we choose $N=200$, $R=80$, $\lambda=0.01$, $p_a=0.05$, $\alpha=3$, $L=28$, $M=24$ and $\delta^2=\frac{R^{-\alpha}}{10}$.

Fig.~\ref{fig:performance_comparison} plots the error probability versus the density of interfering devices $\lambda$, the length of pilot sequences $L$ and the number of antennas $M$. From Fig.~\ref{fig:performance_comparison}, we observe that the optimization-based designs (i.e., proposed joint ML, proposed joint MAP and ML) significantly outperform AMP at the cost of computational complexity increase; the proposed joint ML estimation outperforms ML, especially in the high interference regime; and the proposed joint MAP estimation can reduce the error probability by nearly a half, compared to the proposed joint ML estimation. Note that the performance gain of the proposed joint ML estimation over ML comes from the consideration of interference, and the performance gain of the proposed joint MAP estimation over the proposed joint ML estimation comes from the incorporation of prior knowledge of interference powers and device activities. 
Specifically, from Fig.~\ref{fig:performance_comparison}~(a), we can see that the error probability of each design increases with the density of interfering devices, demonstrating the impact of interference in device activity detection. In addition, we can see that the gap between the proposed joint MAP estimation and the proposed joint ML estimation increases with $\lambda$, which shows that the value of prior knowledge of interference powers increases with their strengths. From Fig.~\ref{fig:performance_comparison}~(b) and (c), we observe that the error probability of each design decreases with $L$ and $M$;  and the gap between the proposed joint MAP estimation and the proposed joint ML estimation increases as $L$ and $M$ decrease, which highlights the benefit of prior information
at small $L$ and $M$.

\section{Conclusion}

In this paper, we considered device activity detection in grant-free random access in a large-scale network with interfering devices. We formulated the problems for the joint ML estimation and joint MAP estimation of both device activities and interference powers, which are DC programming problems. For each problem, we proposed a coordinate descent algorithm to obtain a stationary point. Numerical results demonstrated the importance of explicit consideration of interference and the value of prior information in device activity detection. To our knowledge, this is the first time that techniques from stochastic geometry, estimation and optimization are jointly utilized in device activity detection. Furthermore, this is the first work that explicitly considers the impact of interference generated by massive IoT devices on device activity detection.



\end{document}